\documentclass[]{spie}  

\usepackage{amsmath,amsfonts,amssymb}
\usepackage{graphicx}
\usepackage[colorlinks=true, allcolors=blue]{hyperref}

\title{Visible extreme adaptive optics for GMagAO-X with the triple-stage AO architecture (TSAO).}

\author[a]{Sebastiaan Y. Haffert}
\author[a]{Jared R. Males}
\author[a]{Laird M. Close}
\author[a,b]{Olivier Guyon}
\author[b]{Alexander Hedglen}
\author[b]{Maggie Kautz}
\affil[a]{University of Arizona, Steward Observatory, Tucson, Arizona, United States}
\affil[b]{Wyant College of Optical Science, University of Arizona, Tucson, Arizona, United States}

\authorinfo{Send correspondence to S. Y. Haffert at shaffert@arizona.edu}

\begin{document} 
\maketitle

\begin{abstract}
The Extremely Large Telescopes will require hundreds of actuators across the pupil for high Strehl in the visible. We envision a triple-stage AO (TSAO) system for GMT/GMagAO-X to achieve this. The first stage is a 4K DM controlled by an IR pyramid wavefront sensor that provides the first order correction. The second stage contains the high-order parallel DM of GMagAO-X that has 21000 actuators and contains an interferometric delay line for phasing of each mirror segment. This stage uses a Zernike wavefront sensor for high-order modes and a Holographic Dispersed Fringe Sensor for segment piston control. Finally, the third stage uses a dedicated 3K dm for non-common path aberration control and the coronagraphic wavefront control by using focal plane wavefront sensing and control. The triple stage architecture has been chosen to create simpler decoupled control loops. This work describes the performance of the proposed triple-stage AO architecture for ExAO with GMagAO-X.
\end{abstract}

\keywords{Adaptive optics, Exoplanets, Wavefront sensing, Wavefront control, coronagraphy, Giant Magellan Telescope}

\section{INTRODUCTION}
\label{sec:intro}
Direct imaging of exoplanets is a challenging endeavor. Especially if we are using ground-based telescopes. However, detecting Earth-like planets, and hopefully even habitable planets, is the primary science driver for the next generation of extremely large telescopes (ELTs). The ELTs will be build on Earth, and will have to image through the Earth's atmosphere. The turbulence in the atmosphere degrades the spatial resolution of the telescopes and limits their resolution to an arcsecond. A second challenge is the large contrast between the exoplanet and its host star. Earth-like planets are easily tens of billions times fainter at separation of a couple times the diffraction limit of the telescopes. Imaging such exoplanets will require extreme high performance adaptive optics systems.

Most of the current direct imaging instruments have been designed to operate at near-infrared wavelengths (Y to L band). This was mainly a technical limitation. Longer wavelengths need less precise correction to achieve similar performance as compared to short wavelengths. Therefore, it was easier to start at longer wavelengths for the first generation of exoplanet imagers such as VLT/NACO\cite{lenzen2003naos, rousset2003naos}. This made such instruments suited to hunt for young hot giant gas planets. These planets still retain a large amount of the heat that built up during their formation phase. The heat makes the planets self-luminous in the infrared and therefore bright at infrared wavelengths. Such planets usually are at contrast levels of only $10^{-4}$ to $10^{-6}$. The first generation of HCI instruments lead to the development of instruments that were finely tuned for exoplanet imaging.  The second generation of HCI instruments such as SPHERE, GPI and scexao focused on improving the contrast within the control radius. This was done by adding more advanced coronagraphs and wavefront control. The success of these instruments lead to deeper sensitivities and the discovery of more planets and brown dwarfs.

The Magellan Adaptive Optics system eXtreme (MagAO-X) has been specifically designed to push extreme AO towards optical wavelengths\cite{males2022magaox}. There is exciting exoplanets science at such short wavelengths: planets in reflected light\cite{serindag2019testing}, detection of bio-signatures in atmospheres \cite{schwieterman2018exoplanet} and accreting proto-planets\cite{haffert2019two}. Recent commissioning observations have shown that MagAO-X is close to its designed specification\cite{males2022magaox}. MagAO-X will start to tackle the optical science cases. However, only a handful targets are available at the angular resolution and sensitivity of MagAO-X. We need a system with similar specifications to MagAO-X on the ELTs to survey many targets to get a broad understanding of exoplanets.

The Giant Magellan Telescope Adaptive Optics Extreme\cite{males2022gmagaox} (GMagAO-X) instrument is a direct imaging instrument under development at the UofA for the 24.5-meter Giant Magellan Telescope (GMT). The University of Arizona Space Institute (UASI) funded a conceptual design study for GMagAO-X beginning in February 2021. A positive Conceptual Design Review (CoDR) on September 14th, 2021, determined that the project could move into a UASI-funded preliminary design phase. GMagAO-X is currently the only direct imaging instrument under development for any of the large ground-based telescopes (GMT, Thirty Meter Telescope (TMT) and the European Extremely Large Telescope (E-ELT)) that can observe exoplanets in the visible spectral range. We will be able to image and characterize several hundred of planets with GMagAO-X. GMagAO-X will use a 21000 actuator deformable mirror (DM) to achieve high sensitivity at visible wavelengths. The 21000-actuator DM will use novel pupil slicing and recombining techniques to implement a parallel DM where each GMT segment will get its own 3K Boston Micromachine DM\cite{close2022gmagaoxDM}. 

A major challenge will be the wavefront sensing and control architecture for such a large actuator DM. Current cameras put strong limitations on the number of available pixels and read-out speeds. A strong driver in the design of GMagAO-X is to use hardware components that are available today to make sure we can build the instrument whenever the instrument gets greenlighted. This proceeding will describe an approach that is being investigated for GMagAO-X that works with current hardware. In Section 2 we will describe the overall architecture. Each of the sub-components of the full loop are described in Sections 3, 4 and 5.

\section{OVERVIEW OF THE ARCHITECTURE}
The main challenge for the AO system of GMagAO-X is the large amount of pixels that are required to sample the pupil. The 21000 actuator DM has a projected pitch of 14 cm onto the primary mirrors of the GMT. An equivalent monolithic DM would have 175x175 actuators across the pupil. We follow the same approach as MagAO-X which increased its pupil sampling by 25\% to lower aliasing errors. This means we have to sample the GMT pupil with $175 \times 1.25\approx220$ pixels across the full aperture (over all segments). Most of the new AO system use either the OCAM 2K (240x240 pixels) for wavefront sensing in the visible or the CRED ONE (320 x 256) for wavefront sensing in the near-infrared. These sensors are cameras that have Electron Multiplying (EM) gain capability which reduces the read noise to below 1 electron rms. Low read noise detectors are necessary as this is a major limitation for the low-flux performance of AO systems. While larger sensors are in development, GMagAO-X follows the design philosophy that we will use currently available hardware to make us ready to build the system as soon as possible. This does not mean that new technology will not be considered for GMagAO-X, but that we do not want to be dependent on the development future technology.

If we limit our available pixels to 240 pixels (the lowest denominator of the current high-speed EMCCDs) then that is already close to the required sampling of a single pupil. Any wavefront sensor that needs multiple pupil images can not be used with a single detector. One of the first designs for GMagAO-X was to use either a four-sided or three-sided pyramid wavefront sensor\cite{ragazzoni1996pupil,schatz2021three} where each pupil is send to its own camera. While this might solve the pixel issue, it comes with a lot of synchronization and software challenges. Single pupil wavefront sensors are interferometric wavefronts sensors. Starlight is generally temporally incoherent because we use broadband light. This precludes the use of heterodyne-like interferometry measurements. The starlight itself has to create its own reference beam. Going even further, the interferometer needs to be a common path interferometer. If the reference beam is created in a separate optical path any differential aberration (most likely vibrations) will lower the performance. From all common path interferometers it is the Zernike wavefront sensor (ZWFS) that is the most efficient in its use of light \cite{zernike1942phase,n2013calibration}.

The major downside of any interferometer is the dynamic range. The dynamic range of a ZWFS is significantly smaller than what is required to sense open-loop atmospheric turbulence. This is also the reason why the ZWFS has only been used as a second stage sensor \cite{n2013calibration}. We propose to use a multi-stage system where a first stage AO system is used to take care of the large low-order aberrations. The HODM has a stroke requirement of 3.5 um which is not enough to capture the full range of turbulence. GMagAO-X was always envisioned to operate with a Woofer-Tweeter architecture similar to that of MagAO-X. Here we propose to add a separate low-order wavefront sensor to directly control the woofer (e.g. the low-order DM (LODM)) to create the first stage AO loop. This cleans up the wavefront aberrations enough that we can use a ZWFS for the high-order loop that controls the 21-kilo DM. A second part of the high-order loop contains a differential piston sensor. Sensing differential piston is possible with the ZWFS, but because the ZWFS is an interferometric wavefront sensor it is sensitive to phase wrapping effects. Therefore, we include the Holographic dispersed fringe sensor (HDFS) as a piston sensor for the GMT \cite{haffert2022phasing}. This sensor has been demonstrated to reach <30 nm piston rms for the GMT in a lab environment \cite{haffert2022phasing,hedglen2022lab}. We assume that the HDFS is tracking and controlling the differential piston for the simulations in this work. This is implemented by removing all differential piston from the input disturbances. Later work will include the HDFS in the end-to-end simulations.

The low-order AO loop will control the monolithic 64x64 actuator woofer DM. We are currently baselining an ALPAO 3228 DM beacuse of its 15 um stroke. The low-order loop will only need $62\times 1.25\approx80$ pixels across the full pupil. Such a pupil easily fits multiple times on either the OCAM2K or the CRED ONE. We are considering either a three or four-sided pyramid wavefront sensor to control the LODM. In this proceeding we explore the four-sided PWFS. However, the three-sided PWFS has been shown to have equivalent performance for bright objects\cite{schatz2021three} and superior performance for faint objects\cite{codona2018comparative}.

\begin{figure*}[htbp]
 \centering
 \includegraphics[width=\textwidth]{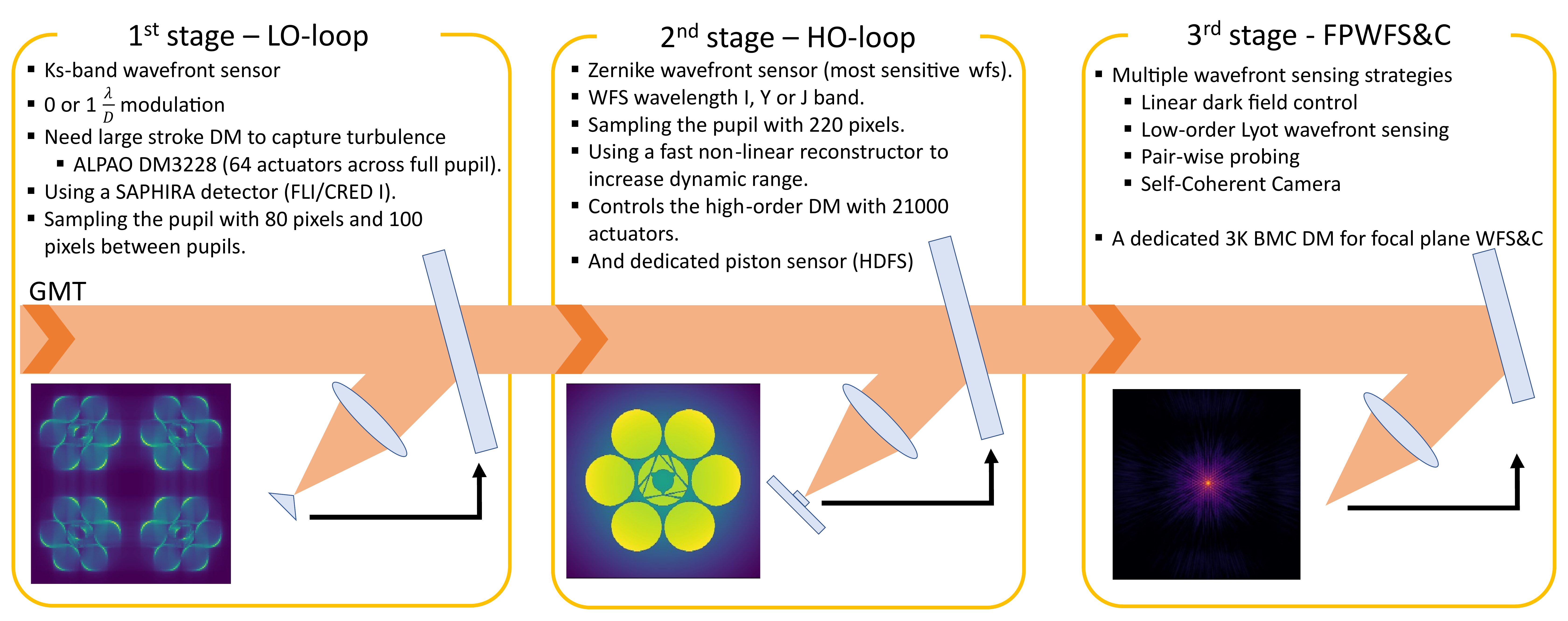}
 \caption{An overview of the triple stage AO system that is envisioned for GMagAO-X. It starts with a low-order loop that has a pyramid wavefront sensor driving the 3228 actuator woofer. A second stage with a Zernike wavefront sensor and the high-order 21K DM is used to achieve very high Strehl. The well corrected wavefront is then send into the coronagraph that has multiple wavefront sensors to control non-common path aberrations and create dark holes around the star.}
 \label{fig:overview}
\end{figure*}

The LO loop and HO loop together will deliver PSFs with high Strehl which are send to the coronagraph. Any non-common path aberration (NCPA) between the AO loops and the coronagraph optics will create stellar leakage around the coronagraph. These NCPA will have to be controlled and removed. GMagAO-X will contain several focal-plane wavefront sensors (FPWFS) and coronagraphic wavefront sensors (CWS) to measure the NCPA and create dark holes around the PSF. The main problem then comes to control of the aberrations. The NCPA could in principle be off-loaded to the HO loop. However, operating the ZWFS (or even the PWFS for that matter) around a non-zero point will lead to non-linear interactions that are difficult to track and calibrate. It is easier if the responsibilities of the AO system and the coronagraph are separated. This can be achieved by adding a dedicated DM in the coronagraph that is purely used for NCPA correction and dark hole digging. This is the approach that is implemented on MagAO-X \cite{males2022magaox}. Separating the responsibilities also makes the control loop design much easier. We do not have to take into account any dynamic interaction between the coronagraph and the AO system. All together, GMagAO-X will have three stages in its wavefront control architecture. An overview of the proposed multi-stage AO system can be seen in Figure \ref{fig:overview}.

\section{THE LOW-ORDER AO LOOP}
The LO loop consists of the 3000 actuator DM and a variant on the pyramid wavefront sensor. One of the major design choices for the low-order loop is to include a dedicated LODM inside GMagAO-X. GMagAO-X will not use the Adaptive Secondary Mirrors (ASM) of the GMT. This is mainly driven by ease of calibration. It is tricky to calibrate the ASM without complicated illumination optics, while it is much easier for an internal DM.

\begin{figure*}[htbp]
 \centering
 \includegraphics[width=\textwidth]{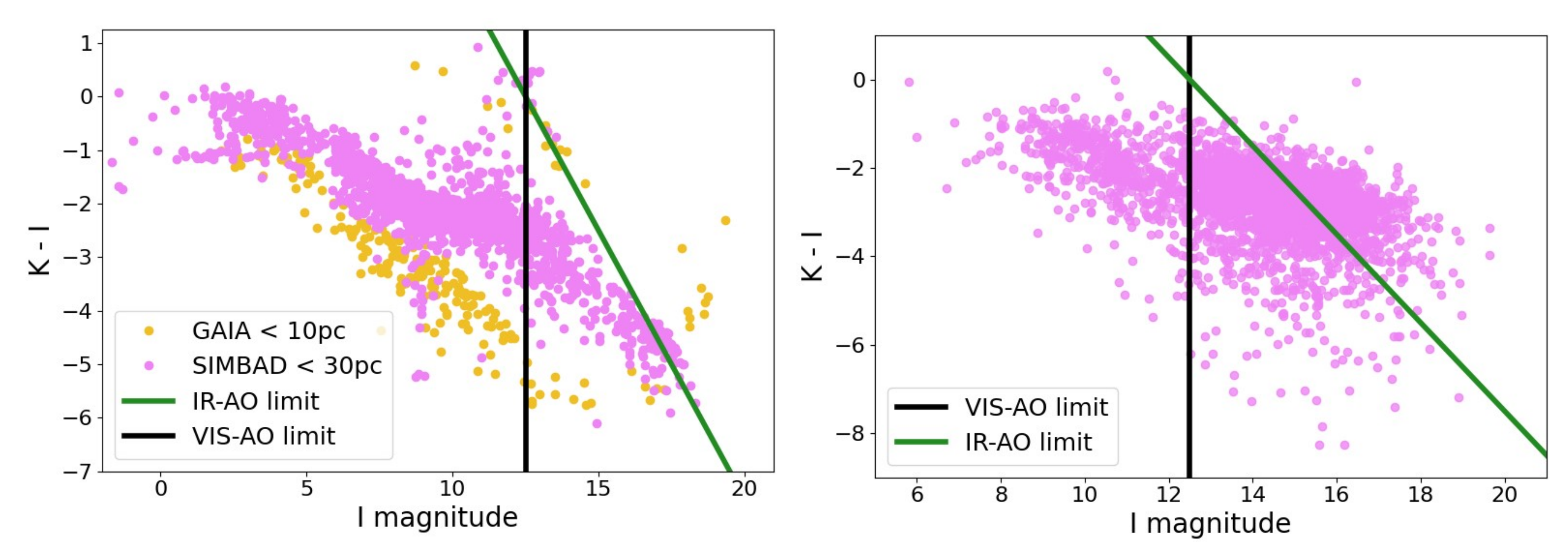}
 \caption{The figure on the left shows a representative sample of close by stars for GMagAO-X science objective 1 and 2.  While the figure on the right shows all young T-Tauri stars for science objective 3. Both figures show that a K-band wavefront sensor would enable science on many more targets than a visible wavefront sensor. And, most T-Tauri stars are much brighter in K-band which leads to higher Strehl.}
 \label{fig:guidestars}
\end{figure*}

One of the main choices we have to consider is the wavelength that will be used to drive the LO loop. MagAO-X uses I-band photons for wavefront sensing with its PWFS. That wavelength band provides the best sensitivity for M stars \cite{guyon2018wavefront}, which are the prime targets of GMagAO-X. However, the short wavelength of 0.8 $\mu$m means that we have to modulate the PWFS to increase its capture ranges. The PWFS at MagAO-X is often modulated by 3 $\lambda/D$ as this provides a stable system that is robust in most seeing conditions. The modulation lowers the sensitivity of the wavefront sensor significantly\cite{fauvarque2017general}. While visible wavefront sensing like I-band is the most common band, more and more systems are considering near-infrared wavefront sensors \cite{bond2018adaptive,boccaletti2020sphere}. This is driven by both the amount of targets and the non-linearity effects of the PWFS. The amount of non-linearity is set by the wavelength. Sensing the wavefront at longer wavelengths makes the PWFS more linear\cite{bond2018adaptive}. Most importantly, the amount of targets increases substantially. Figure \ref{fig:guidestars} shows the amount of targets that are available for GMagAO-X if we use an I-band wavefront sensor or a K-band wavefront sensor. There are two important science cases:
\begin{itemize}
  \item Exoplanets in reflected light. These companions need to orbit their host star at relatively close orbits to capture and reflect enough light to make them detectable\cite{guyon2018wavefront}. The sample of targets are all the stars that are in the Solar Neighborhood. We made a selection by looking for all objects within 10 pc within in the GAIA catalogue and all objects within 30 pc in the SIMBAD database.
  \item Accreting proto-planets. These planets are still young and embedded in a (gapped) circum-stellar disk. Such stars are usually classified as T-Tauri stars. Our second sample contains all T-Tauri stars that are visible from the southern hemisphere.
\end{itemize}
The amount of available targets increases by a factor 3 if we use a K-band wavefront sensor instead of an I-band wavefront sensor. The loss in sensitivity in K-band is compensated by the additional amount of targets that are observable. All most all T-Tauri stars are significantly brighter in K-band, which will partially compensate for the loss in sensitivity. Another driver for the choice for K-band is that GMagAO-X will support science observations from R-band to H-band. All of the K-band photons can be used for wavefront sensing. This will increase the throughput to the science channels because no photons have to be shared.

\begin{figure*}[htbp]
 \centering
 \includegraphics[width=0.7\textwidth]{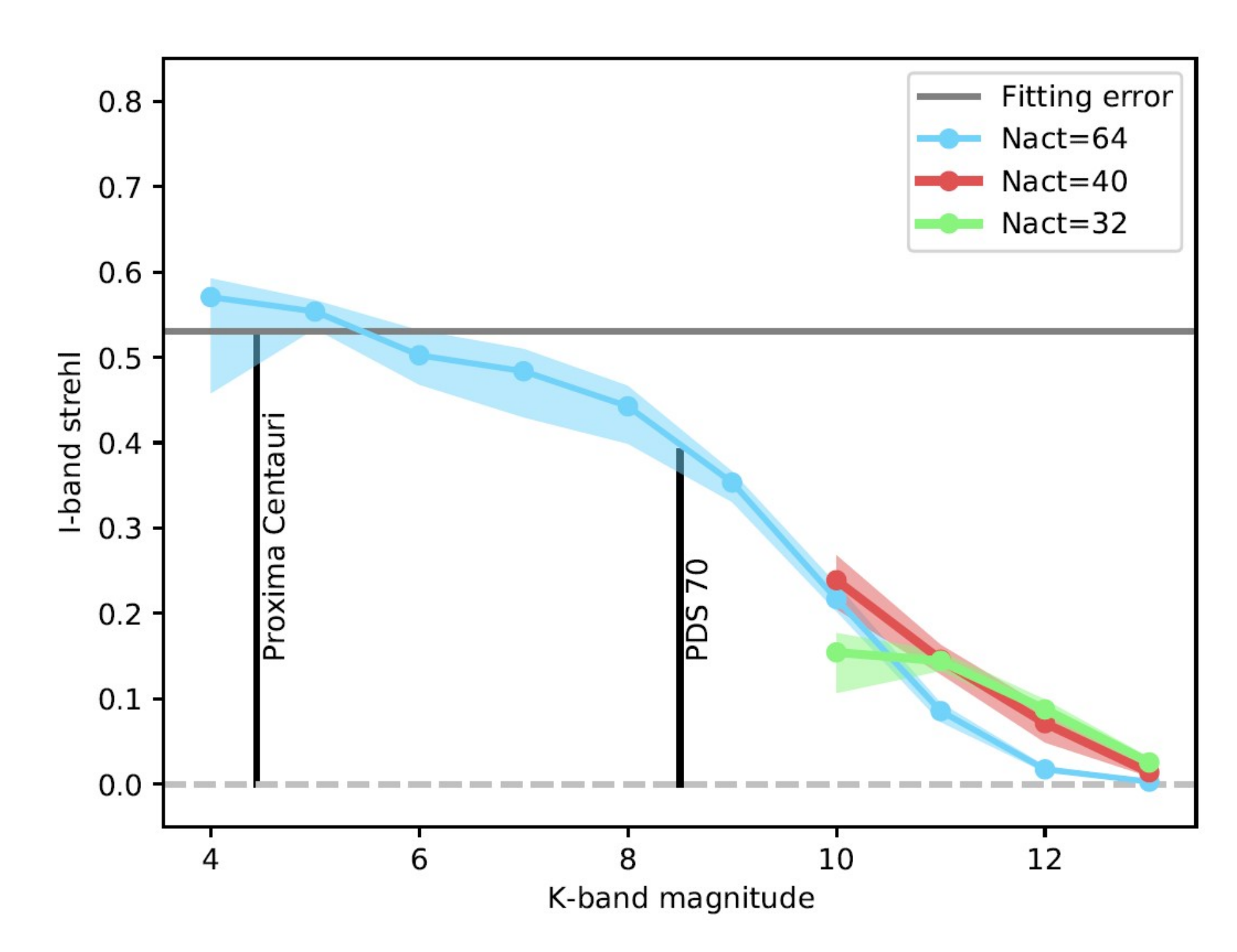}
 \caption{The Strehl in I-band after the first stage as function of K-band magnitude. A higher Strehl can be achieved on fainter target if we control less modes. This has been explored by varying the number of controlled modes. The first stage determines the guide star magnitude limit because for faints stars the high-order DM is not necessary. The first stage achieves $>5$\% Strehl in I-band for K=12.}
 \label{fig:firststage}
\end{figure*}

The K-band LO-loop performance is predicted by end-to-end simulations. The parameters of the LO system can be found in Table \ref{tab:loe2esims}. The LO loop uses a simple leaky integrator for the controller, where the gain is optimized for each guide star magnitude. We do not expect significant optical gain effects because the sensing happens at K-band. Therefore, we have not applied any optical gain corrections for these simulations. We ran 5 trials of 500 iterations for each combination of guidestar magnitude and gain. The median Strehl across the last 250 iterations of all 5 trials is shown for each magnitude in Figure \ref{fig:firststage}. The band around the curve are the 16\% and 84\% quantiles of the Strehl.
\begin{table}[]
\centering
\caption{The parameters of the end-to-end simulations of the LO loop.}
\begin{tabular}{l|l|l}
\hline
Parameter &  Value & Unit  \\ \hline\hline
$\lambda_0$ & 2.2 & $\mu$m \\ \hline
$\Delta \lambda / \lambda$ & 0.2 &  \\ \hline
read-noise & 0.3 & $e^{-}$/pixel \\ \hline
background & 12.5 & mag arcsecond$^{-2}$ \\ \hline
loop speed & 1.0 & kHz \\ \hline
number of modes & 3200 & kHz \\ \hline
modulation & 1 & $\lambda/D$ \\ \hline
pupil sampling & 80 & pixels \\ \hline
pupil distance & 100 & pixels \\ \hline
\end{tabular}
\label{tab:loe2esims}
\end{table}

There are not enough photons to control all modes for the faintest targets. We also ran the same gain optimization procedure for magnitudes 10 to 13 with 800 modes and with 1200 modes. We saw a significant increase in performance at 11, 12 and 13th magnitude. The system still delivers 10\% Strehl at 12th magnitude in I-band.

\section{THE HIGH-ORDER AO LOOP}
The high-order loop uses a ZWFS to control the high-order DM. The phase reconstruction is still a challenge even with the reduced number of pixels that ZWFS uses. The ZWFS can in principle use a linear reconstructor in the small phase regime\cite{n2013calibration}. Most AO systems would use a straightforward Matrix-Vector Multiplication (MVM) for that. This matrix has a size of 21000 by 35000 elements. Calibrating this matrix with a push-pull approach would take a significant amount of time. Therefore, we are reconstructing the input phase directly. Mathematically the ZWFS can be described as,
\begin{equation}
    E_{o} = E_{i} + m*(e^{i\delta}-1)E_{i}.
\end{equation}
Here, $E_{o}$ is the output electric field and $E_{i}$ the input electric field. The ZWFS applies a low-pass filter through the phase dot (with step size $\delta$) on the input electric field. This is described by the convolution of $E_{i}$ with the pupil response of the ZWFS mask $m$. For a well corrected beam the spatially filtered electric field is very close to the diffraction-limited electric field \cite{n2013calibration,steeves2020picometer}. The filtered part can be replaced by its diffraction-limited counterpart with a scalar correction for the Strehl loss. This effectively means we are interfering the aberrated electric field with a spatially filtered version of the diffraction-limited beam. This is a simple two beam interference problem,
\begin{equation}
    E_o = E_i + E_r.
\end{equation}
In this equation, $m*(e^{i\delta}-1)E_{i}$ has been replaced by the modelled reference electric field $E_r$. The intensity of the measurement is,
\begin{equation}
    I_o = I_i + I_r + 2|E_i||E_r|\cos{\left(\phi_i - \phi_r\right)}.
\end{equation}
The input intensity, $I_i$, is just the measured pupil intensity and does not depend on the phase. This can be measured and subtracted. Similarly, the reference intensity can also be subtracted. The phase reconstruction is,
\begin{equation}
    \phi_i = \arccos{\left[\frac{I_o - I_i - I_r}{2\sqrt{I_iI_r}}\right]} + \phi_r.
\end{equation}
This reconstructor takes all non-linearities of the ZWFS into account, if the reference field has been corrected for the Strehl loss. The direct phase reconstruction is much quicker than a MVM multiplication. The actuator voltages signals are calculated by projecting the reconstructed phase onto the voltages. This is done through a sparse MVM because the actuators have a very localized influence function. Therefore, it is faster to project the phase onto the actuators with a sparse MVM.

\begin{figure*}[htbp]
 \centering
 \includegraphics[width=\textwidth]{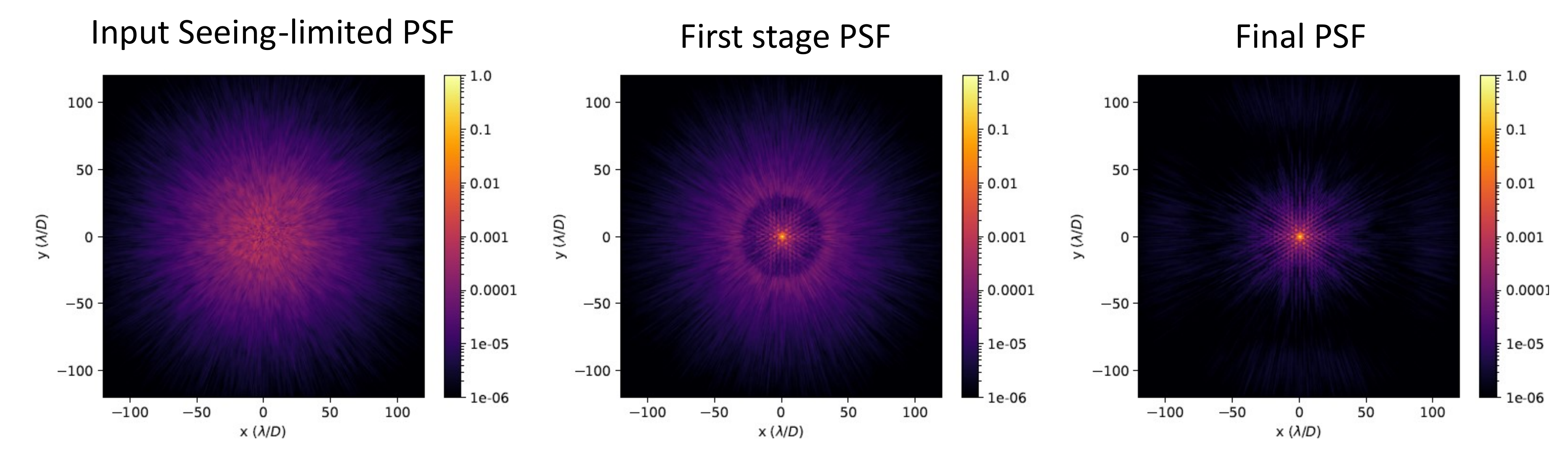}
 \caption{A representative PSF for a bright target. The image was simulated with a 20\% bandpass centered on 1 um. The final PSF has a Strehl of above 95\% in median seeing condition. The proposed architecture can deliver the required performance in median conditions.}
 \label{fig:expectedpsf}
\end{figure*}

\begin{figure*}[htbp]
 \centering
 \includegraphics[width=\textwidth]{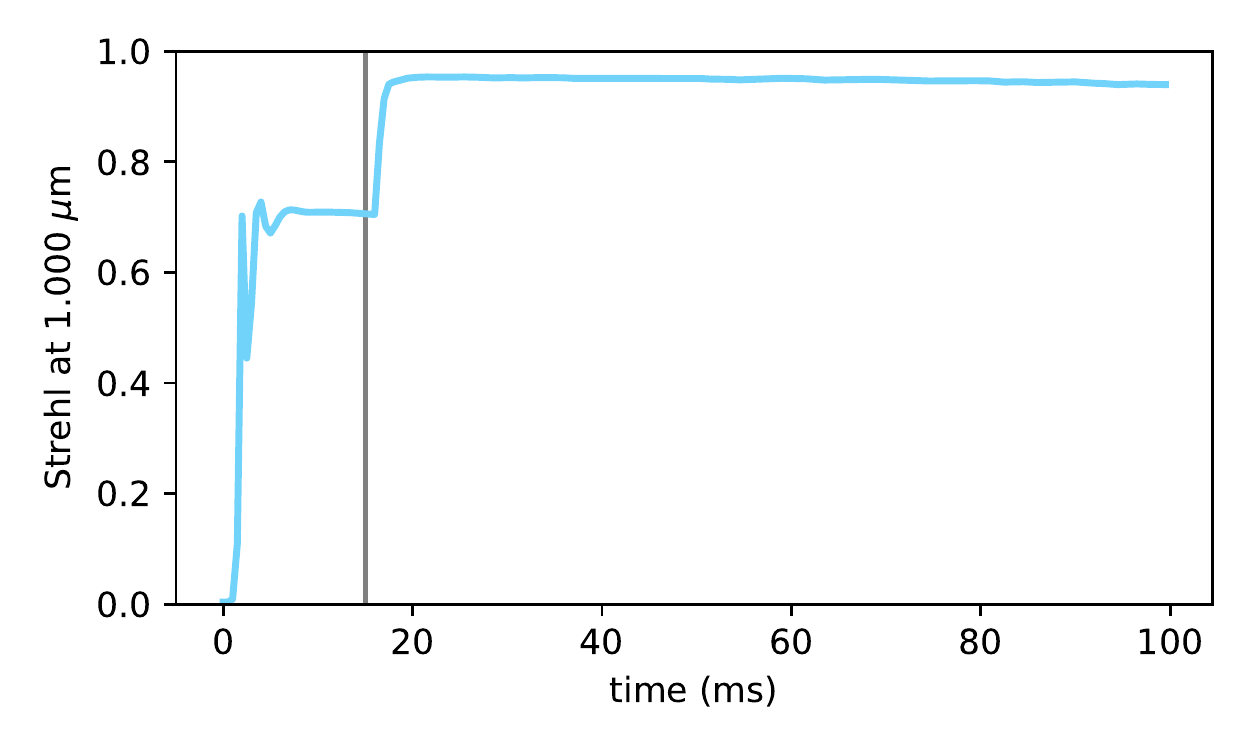}
 \caption{The figure on the left shows a representative sample of close by stars for GMagAO-X science objective 1 and 2.  While the figure on the right shows all young T-Tauri stars for science objective 3. Both figures show that a K-band wavefront sensor would enable science on many more targets than a visible wavefront sensor. And, most T-Tauri stars are much brighter in K-band which leads to higher Strehl.}
 \label{fig:msstrehl}
\end{figure*}

The HO loop is closed after 15 iterations of the LO loop in the end-to-end simulations. In practice this will most likely take longer if human interaction is necessary to close both loops. We use a 0th magnitude guidestar for our first end-to-end simulation of the multi-stage AO system (MSAO). For this guidestar magnitude both the LO and HO loop run at 2 kHz with optimized gains. The ZWFS uses J-band light for these simulations. The PSFs at various stages during the loop are shown in Figure \ref{fig:expectedpsf}. The seeing halo is completely removed up to 90 $\lambda/D$ after the second stage. The Strehl reaches nearly 95\% at $1\mu$m which can be seen in Figure \ref{fig:msstrehl}. This shows that this system can reach the required performance for high-contrast imaging at short wavelengths.

\section{CONCLUSION AND OUTLOOK}
This proceeding describes the current wavefront sensing and control architecture for GMagAO-X. A system with multiple stages is considered because of the design philosophy. The system will use a first stage K-band pyramid wavefront sensor running at 1 or 2 kHz. Followed by a Zernike wavefront sensor at bluer wavelengths to control the high-order loop. This combination of wavefront sensors allows us to control our 21000 actuator DM and achieve a Strehl of 95\% at 1$\mu$m on the GMT.
The current simulations are the first end-to-end simulations of GMagAO-X and do not include all sources of noise and all input disturbances. Therefore, our next step will be to expand the simulations to make them more representative:

\begin{itemize}
  \item The input disturbance will need to include vibrations caused by the telescope and windshake.
  \item Include differential piston, which we have now neglected.
  \item More representative detector noise.
  \item Advanced control algorithms such a predictive control\cite{haffert2021data} and optical gain compensation.
\end{itemize}

\acknowledgments     
Support for this work was provided by NASA through the NASA Hubble Fellowship grant \#HST-HF2-51436.001-A awarded by the Space Telescope Science Institute, which is operated by the Association of Universities for Research in Astronomy, Incorporated, under NASA contract NAS5-26555. 

\bibliography{report} 
\bibliographystyle{spiebib} 

\end{document}